\documentclass[runningheads]{llncs}
\usepackage[T1]{fontenc}
\usepackage{graphicx}
\usepackage{tabularx}
\usepackage{amsmath}
\usepackage{hyperref}
\usepackage{graphicx}
\usepackage{authblk}
\usepackage{multirow}
\begin{document}
\title{The ethical situation of DALL-E 2}
%

\author{Eduard Florin Hogea \inst{1,\dagger}}
\author{Josep Maria Rocafort Ferrer \inst{2, \dagger}}

\authorrunning{E. Hogea \and
J. Rocafort}

\institute{
\email{eduard.hogea00@e-uvt.ro}\\ \and
\email{josem.rocafortf@gmail.com}
}
\affil{$\dagger$ Computer Science Department, UAB, Catalonia, Spain}

%
\maketitle              
\begin{abstract}
A hot topic of Artificial Intelligence right now is image generation from prompts. DALL-E 2 is one of the biggest names in this domain, as it allows people to create images from simple text inputs, to even more complicated ones. The company that made this possible, OpenAI, has assured everyone that visited their website that “Our mission is to ensure that artificial general intelligence benefits all humanity”. A noble idea in our opinion, that also stood as the motive behind us choosing this subject. This paper analyzes the ethical implications of an AI image generative system, with an emphasis on how society is responding to it, how it probably will and how it should if all the right measures are taken.

\end{abstract}
\section{Introduction}
The field of Artificial Intelligence has exploded in popularity thanks to significant strides in recent years in machine learning and deep learning techniques. These advancements have led to the development of powerful AI systems, which is the focus of this paper. One such system is DALL-E 2 \cite{two}, a state-of-the-art AI language generation model developed by OpenAI, which has interesting ethical considerations. DALL-E 2 can generate images, music, and even code from natural language prompts, potentially revolutionizing various industries, from media and entertainment to healthcare and education.

However, with the increasing power and capabilities of AI systems like DALL-E 2, it is essential to consider the ethical implications of their use. Ethical considerations are important to ensure that AI systems are developed and deployed responsibly and benefit society as a whole. Failure to consider these ethical implications can lead to unintended consequences that may harm individuals or groups and undermine public trust in AI. This paper will specifically address these concerns and illustrate how unchecked support for AI systems can harm public opinion.

The ethical challenges posed by AI systems like DALL-E 2 are not isolated. Our previous research has addressed similar issues in different AI applications. For instance, in \cite{onchis2023neuro}, we explored a neuro-symbolic model for damage detection in cantilever beams, highlighting the importance of robust AI models in industrial applications. Additionally, in \cite{hogea2024fetril++}, we proposed FeTrIL++, a method for exemplar-free class-incremental learning, showcasing the evolution of AI in learning and adaptation without compromising previous knowledge.

This image generator is based on Deep Learning, requiring a massive dataset for training the model. The dataset images were pulled from the internet, raising ethical concerns about copyright issues and accusations of DALL-E "stealing" art styles. This architecture's application in creative industries democratizes access to creative tools but also raises questions about authorship, ownership, and royalties. Similar ethical concerns are discussed in my work on monitoring security alerts using a neuro-symbolic classifier \cite{onchis2022neuro}, where we emphasized the need for optimized and ethical AI deployments.

Moreover, AI's potential for misuse, such as creating deep fakes, is another critical issue. We will briefly discuss these issues in the following sections, presenting the system's architecture and the creators' efforts to mitigate misuse. Our ongoing research, including predictive modeling for diabetes using GraphLIME \cite{costi2024predictive}, continues to explore the balance between innovation and ethical responsibility in AI development.

\section{Understanding what can DALL-E 2 actually do}
We would like to start this part of the project with some images that were generated from such systems.

\begin{figure}[!htbp]
    \centering
    \includegraphics[scale = 0.25]{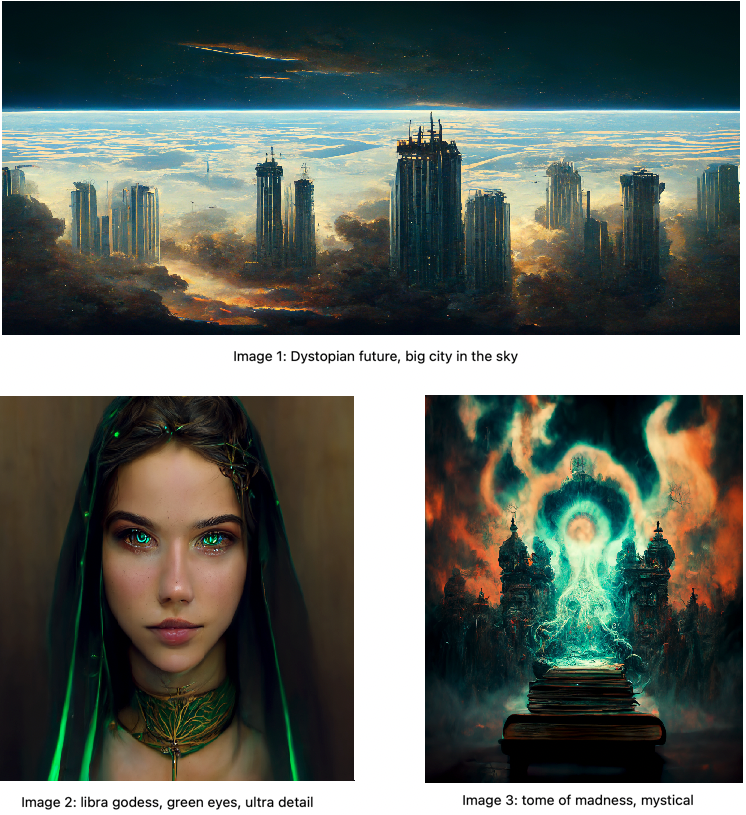}
    \caption{Images generated by us with a similar system called Midjourney AI, with the prompts shown below each of them. }
    \label{fig:my_label}
\end{figure}

Simply put, we think they are quite fascinating. But we understand that by being this good, it can be detrimental to the system itself. DALL-E 2’s incredible ability at generating hyperrealistic images has also led to criticism about possible usages. One could use it to massively create fake news and deep-fakes of politicians and celebrities, creating a chaotic internet where it would be nearly impossible to detect if something was real or not. Also, it could be used to impersonate a person in a porn image/video that never was involved in any of that, it would have obvious bad implications and affections on that person's life.

If we consider what the impact of DALL-E 2 or other alike AIs and think about their post-mortem, and what could go very wrong with them, another thing that comes to mind is the workload that is reduced on artists and graphic designers. The ethical part is, similarly to the point made about authorship before, how people's jobs are going to be affected by this AI? Raising this ethical issue, it is important to consider the implications of such developments for workers and to ensure that measures are put in place to mitigate any negative effects on employment. There are, however, also the philosophical implications about if one image produced by a machine or a human has the same value.

\begin{figure}[!htbp]
    \centering
    \includegraphics[scale = 0.30]{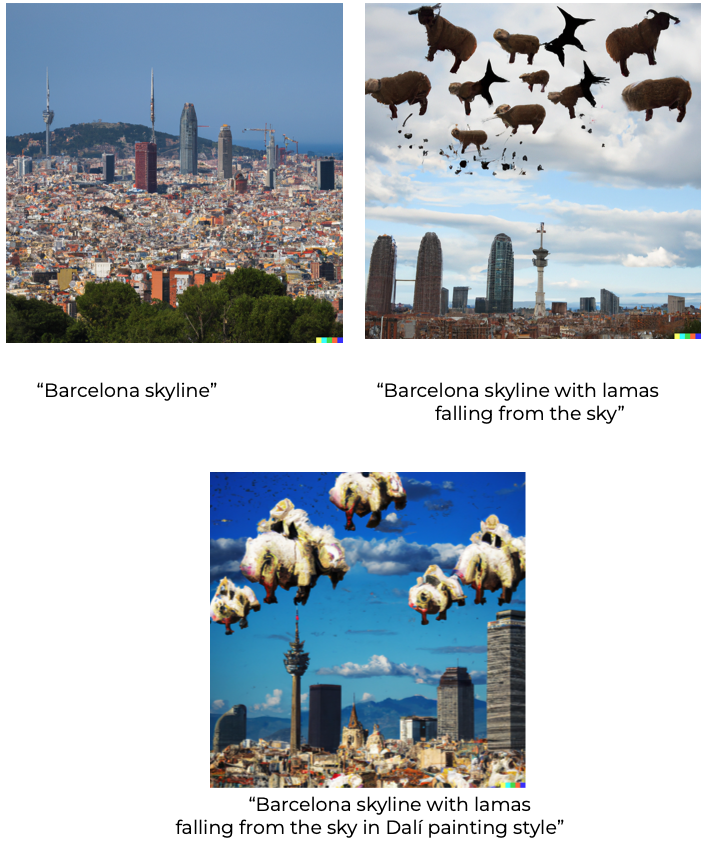}
    \caption{Images generated from the correspondent prompts with DALL-E 2.}
    \label{fig:my_label}
\end{figure}

In Figure 2 we have some examples that we generated using DALL-E 2, and we can observe that the output isn’t perfect, but with the evolution of these AIs and using the right prompts better outputs can be reached. The last one depicting the actual usage of this system in a way that can steal someone’s art style.

Visualizing the system should help better understand the situation. For what, we have the presented map in Figure 3 that is based on the figure provided in \cite{one}, a paper that analyzes DALL-E 2. Two actors involved can be seen. We have the user, that uses a prompt to generate the desired image, on the other hand, we have the already existing classic artists or any kind of digital image creator, including photographers.
Such models need data to be trained on, and these images in some cases are acquired from various artists. The ethical issues aforementioned are based on the training part of this model.

\begin{figure}[!htbp]
    \centering
    \includegraphics[scale = 0.5]{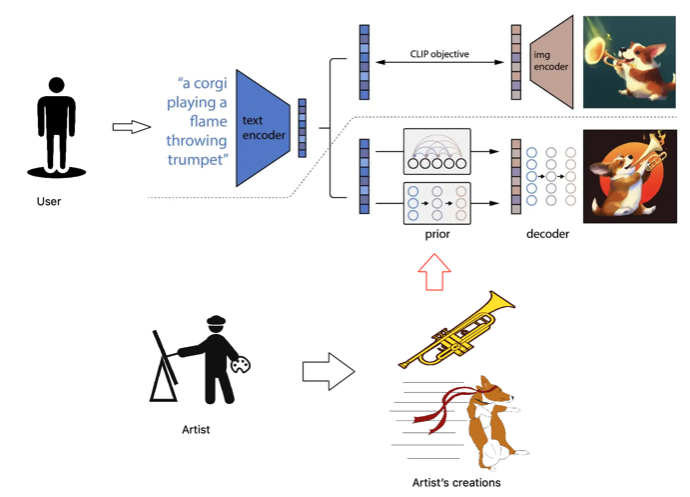}
    \caption{Map of involved actors. The main parts of the system are shown in a flow diagram, with an actual use case where one artist can have two separate images used to create a new one based on the prompt of the user. It can be noted that in the upper part of the presented figure, there are two parts delimited by a dotted line. The upper part, is the CLIP model presented in the paper \cite{ramesh2022hierarchical}, used to generate a set of visual attributes that correspond to the given textual description. The one below, is the Diffusion model that enhances the image prior generated with CLIP to produce the final image. Both of these models are using data gained from the textual input, but also from the artist's images.}
    \label{fig:my_label}
\end{figure}

\section{Current and potential future use of it}
In the present, the use of programs like DALL-E 2 is highly discussed, and different artists have expressed their thoughts about it. We, in this case, consider that the hatred towards this new AI technology is misplaced, as all art in essence is building on concepts. As a muse inspires artists, an analogy can be made on DALL-E 2 and the art pulled from the internet. As unintended usage goes, the system already provides some security measures. The usage of it in propagandistic ways, obscenity and racial ways is also already blocked.

\subsection*{Futures – What could and should be}
The future is unclear, but we can think on some outcomes, considering the actual mass opinion of DALL-E 2.
The ideal future is where artists don’t get their art processed without their approval, and that the AI system has implemented solutions to evade copying, leaving art unique.

The future that could be, can also be a bad one. The massive use and adoption could lead to people stopping valuing original human art and stopping supporting creators, being more and more few of them over time. Having less human visual creativity out there is stopping advancement of art made classically by humans. We could call it classical art. One other bad scenario, is where there are no effective barriers for unwanted usage to become the main source of interest for people, and all the preventive measures would be in vain.

\section{Following the RRI, (Responsible research innovation) principles}

One way that DALL-E and DALL-E 2 might follow the RRI principles is by facilitating “doing good” and contributing to social justice. For example, the technology could be used to create educational materials or to generate images that promote diversity and inclusion. An actual example on how university projects can be improved with an addition of generated images can be seen in \cite{seven}. Additionally, the researchers and developers working on these AI models should have the necessary skills and knowledge to ensure that they are being developed in a responsible and ethical manner. For this last part, we consider the response of the OpenAI team, behind the DALL-E and DALL-E 2 systems, have respectably done their part. Their response can be seen in \cite{eight}, where they explain how constraint have been applied to prevent some ethical issues such reducing graphic and explicit images, effectively removing the ethical possibility of producing deep fakes initially presented. Another important aspect of RRI is ensuring that the outcomes of research are honestly reported. In the case of DALL-E 2, this could involve being transparent about the capabilities and limitations of the technology, as well as any potential ethical concerns that may arise. It may also be significant to consider the potential impact of the technology on ethnic and cultural diversity, as well as its potential impact on the wellbeing and privacy of individuals.

Furthermore, the environmental impact of DALL-E 2 and further technologies should be taken into account. As for training the models, hours, and hours of intense computations in supercomputers are required, with papers such as \cite{patterson2021carbon} addressing this problem, with respect to Deep Neural Networks and carbon emissions (resulting from the energy consumption). The mentioned paper describes also the implications of training GPT-3 models, which are also a part of the DALL-E architecture.  As the model relies on public datasets, the non-uniform distribution of data may result in algorithm biases. This, together with the constraints of non-violent and graphic contents, actually made the model more likely to reduce the frequency of women in the images, when no gender was mentioned in the input. The article in \cite{five} addresses this issue more in depth, and it also shows cases where the text input may pose a difficulty in how the algorithm interprets it. This directly correlates to issues from the RRI, such as gender imbalances. A solution, in essence, should be to have a balanced dataset to provide a balanced output. With this purpose, transparency of the dataset it was trained on and with the code it ran should be publicly available, and in our case, it is not. An example of a dataset used publicly available, that could be used in developing further similar algorithms is \cite{four}.

The language processing part of the system has been openly discussed, as in the paper \cite{tamkin2021understanding}, with focus on the following:
\begin{itemize}
    \item What are the technical capabilities and limitations of large language models? 
    \item What are the societal effects of widespread use of large language models?
\end{itemize}
This proves that the research on this topic aims to be honest, and that the outcomes of it are made openly available.

Finally, on this part, it is important to address that the research process engages with and involves individuals and different interest groups. This could involve seeking input and feedback from a diverse range of stakeholders, and making the outcomes of the research openly available. By following these RRI principles, DALL-E and DALL-E 2 can be developed and used in a responsible and ethical manner that benefits society as a whole.

\section{Technology and society, a complex relationship}
The social impact of innovations is complex. So complex, that we decided to dedicate a whole part to explain how technology is not good, a bad, but also not a neutral or independent entity. It is, but rather, a deeply intertwined body, with the social, economic, and political factors that shape its development and use. We know this, because we can make analogies from past real scenarios, where at the time state-of-art technologies were proposed, and the effects can be seen just now, one or two decades later. Here are a few examples that we will study, considering the benefits and the deficiencies that emerged with their adoption in the society:

\textbf{Smartphones}: The widespread adoption of smartphones has had a profound impact on society, from how we communicate and access information, to how we navigate and buy things online. However, the development and use of smartphones has also raised ethical concerns, such as the effects on privacy and mental health. In the paper \cite{singh2018impact}, the authors identify some cases where the smartphones had positive and negative effects on students, in regard to focusing on education, psychology, and social aspects. The history of smartphones and their evolution is discussed, as well as statistics on smartphone usage in Malaysia, and a clear take can be seen on the tables that provide a side to side comparison between the benefits and the deficiencies of smartphones. The authors state that the usage can boost the learning experience, the learning knowledge and that it offers the possibility of distance learning. On the other hand, the downsides are almost sevenfold. This research suggests that smartphone usage among university students can have negative effects on their education, relationships, and mental well-being. The constant need to update social media status can distract students from learning and lead to a lack of real-life social interaction. One more thing, the constant use of smartphones during lectures and class can lead to a lack of attention and lower recall of information taught. The overuse of smartphones can also lead to higher levels of depression, anxiety, and a lack of empathy. Moreover, the study suggests that the constant interruption of messages and calls can lead to headaches, interruption in concentration, and difficulty in completing coursework. Furthermore, the research suggests that the constant use of smartphones can develop into addiction, which can negatively impact academic performance and life satisfaction. While the aforementioned paper still saw some good in the relationship between smartphones and society, we have, on a different note, articles such as \cite{davidekova2016digitalization}. In this article, the social impact of smartphones on society and the spread is viewed as a virus. The research findings discuss the negative effects of smartphones, such as addiction, distraction, and negative impact on health. It concludes by providing recommendations to help reduce the negative impact of smartphones on human health.

\textbf{Social media}: Social media platforms have greatly impacted the way we interact and consume information, but also has lead to issues related to misinformation, cyberbullying, and addiction. We found this article \cite{reid2014social}, written by two respectable practitioners in the medical field, to pinpoint, in an intriguing manner, the effect that social media has on teenagers.  95 \% of teens ages 12-17 access the internet, and 70 \% percent do so daily. Social media has become a main form of communication among teens and plays an integral role in their lives. The benefits of social media for adolescents include the ability to connect with friends and strengthen existing relationships. However, the authors also state concern about the negative impact social media may have on adolescent health and development. Aspects of social media such as cyberbullying, suicide, Pro-Self Harm and Pro-Eating Disorder websites and sexting are also all covered in that article. We want to note that the referenced paper is from July 2014 and to make a relevant analogy, we also want to include newer examples.

How is it perceived by society today? To best illustrate this, we would like to highlight the most popular social media application, TikTok. Originated it China in 2016, it saw a worldwide release in 2017 and has quickly become one of the biggest social media platforms with over 1 billion active users and over 2 billion downloads, according to \cite{tik}.

Papers such as \cite{xiuwen2021overview} provide an overview on how TikTok can be used to effectively learn another language, but there are other researchers that are more focused on the bad factors surrounding it. The paper \cite{sha2021research} analyzes the case and correlated the usage of such social media platform to a decrease in attention span and the surge in stress, depression, anxiety, and memory loss in teenagers. Both of the last papers are quite recent (2021) and strengthen the ideas from before that adopting new technologies in society is not a straightforward process.

\textbf{Self-driving cars}: The development of autonomous vehicles has the potential to greatly improve transportation and reduce accidents caused by human error. However, ethical concerns have been raised over how self-driving cars will be programmed to make decisions in dangerous situations and how they will affect employment in the transportation industry. Some of these ethical issues, as presented in \cite{holstein2018ethical} are about safety, security, privacy, trust responsibility and accountability. While it would be a perfect scenario where there was no human error and the cars would be fully automated, what would happen if there was a software problem and the car gets into an accident? Well, as the paper suggests, it is hard to find a solution. It is still regarded as an open problem.

\textbf{Artificial intelligence in healthcare}: AI has the potential to greatly improve healthcare by assisting doctors in diagnostics and treatment, but also raises ethical concerns over data privacy, bias and the human-AI relationship in medical decision-making. There have been known cases that lead to the “AI Winter”, where the expert systems deployed in several fields proved to be too brittle and susceptible to errors. It has regained popularity now, and there are some areas where it excels in performance, but some ethical issues (as in all AI models) concern biases and accountability.

\textbf{The internet}: The internet has revolutionized the way we access information and connect with others, but also has led to issues such as online harassment, hate speech, and the spread of misinformation. The social media and the smartphones inherit all these attributes and by already discussing them, we have a pretty good understanding now what the social impact of internet has.

We wanted to emphasize on this part, as it is a wide known fact that history always repeats itself. We think that the integration of AI generating art systems will behave similarly, and with the detailed explanation provided, we reduced the dimensionality of the social impact of such innovations. It is still a problem, but at least we know what to expect in the near future.

\section{Technological mediation}
With any new technology but even more with something as revolutionary as DALL-E 2 a plan of what impact the technology is going to have on society should be made. In other words it should be planned ahead of time of how we want society to see and use this technology once it is open for the wide-spread use. These are the two main points of what should be managed and supervised during the introduction of this new technology:

\textbf{Embedded values}
Deciding as developers what are the embedded values we want the technology to have once it is part of society set of common tools and culture.
Embedded values refer to the underlying principles and beliefs that are inherent in a technology, and that is certainly shaped by how it is designed, developed, and used.
To control or at least lead in a positive direction, the public opinion during the early adoption has a massive impact. For that, the exposure and media coverage of early use-cases has to be taken into account very carefully. For DALL-E 2 there are two extremes of the same axis that should be avoided, on one hand the hype and the idealization of the technology, believing that it can do everything and is better than humans, we should never let AI be seen as a god or as being superior to humans, human creativity and abilities will always have a place. On the other hand, public opinion should not fall into the demonization of the technology, that ironically also comes from the idealization of these technologies and the fear of complete change. A critic and non polarized approach should be taken so the most positive impact and fully realization of the potential of the technology can be made over time without the harmful effects that come with idealization.

\textbf{Governance and regulation:}
It is very important during development to establishing a clear governance and regulation of the technology. This includes determining who is responsible for overseeing the current development, as well as setting guidelines for how the technology should be used. This can include regulations on data privacy, what can and cannot the model produce (should be controlled from training data) or guidelines for how the technology should be used in certain industries, for example on social media, should it be clearly tagged that an image was generated using AI? There could be implications on issues such as privacy and misinformation. For example, the technology could be used to generate deepfake images, which can be used to spread false information or to impersonate individuals. Therefore, it is crucial to have measures in place to detect and prevent the spread of such images.
This will help ensure that the technology is used in a responsible and ethical manner and that the negative impacts are minimized.

\subsection{Summing up technological mediation}
When introducing a revolutionary technology like DALL-E 2, it is vital to plan and manage the technology's impact on society. This includes deciding the embedded values, identifying potential use-cases, and establishing governance and regulation. By carefully shaping the public opinion during early adoption with a non-polarized approach, developers can ensure that the technology is seen in a positive light. By identifying potential use-cases and assessing their ethical implications, the technology can be developed and implemented in a way that maximizes its positive impact and minimizes negative consequences. Furthermore, by having clear governance and regulation, the technology can be used responsibly and ethically, and negative impacts can be minimized. Additionally, the potential societal impacts of the technology must be taken into account, so that measures can be taken to mitigate any negative effects. By considering these factors, DALL-E 2 can be introduced to society in a way that maximizes its potential for positive impact.

\section{Conclusion}

With this paper, we highlighted the most important ethical issues that DALL-E 2 faces. We have presented the architecture of the system, the potential it has (both in the good cases and in the bad ones), how it is being used right now and what should be changed in the future to prevent its unethical usage.

To strengthen our ideas, we made references to trusted sources and showed how the application of such systems should follow the Responsible Research Innovation principles. With respect to technological mediation and the social impact of this system, we conclude that, introducing a revolutionary technology like DALL-E 2 requires careful planning and management to ensure its positive impact on society. This includes deciding on embedded values, identifying potential use-cases, and establishing governance and regulation. The developers (in this case OpenAI) should aim to shape public opinion during early adoption with a non-polarized approach, and assess the ethical implications of potential use-cases. From our point of view, and as demonstrated by their acts, we think the team working behind the scenes that made this system possible has done an excellent job and their noble ideas should be mentioned clear.

Finally, we think that DALL-E 2 can be successfully introduced to society, without many ethical issues, if the right actions are taken to reduce any negative impact and if all the potential use cases scenarios are taken into account in the future.

%
%
%

\begin{thebibliography}{10}
\bibitem{one}
A.~Pawar, ``A first look at dall-e 2 — how it works under the hood.''
\bibitem{five}
S.~STRICKLAND, ``Dall-e 2’s failures are the most interesting thing about
  it.''
\end{thebibliography}

%





\end{document}